\documentclass[10pt]{article}

\usepackage{amsmath}
\usepackage{amssymb}
\usepackage{graphicx}

\usepackage{cite}

\usepackage{color}

 \usepackage{url}
\usepackage{amsfonts}
 \usepackage{theorem}
\usepackage{graphicx}
\usepackage{enumerate}

 \usepackage{verbatim}
 \usepackage{amssymb}
\usepackage{amsbsy}
 \usepackage{setspace}

 \newcommand{\updt}[1]{{\color{black}#1}}

\newlength\savedwidth

\usepackage[labelfont=bf,labelsep=period,justification=raggedright]{caption}

\makeatletter
\renewcommand{\@biblabel}[1]{\quad#1.}
\makeatother

\theoremstyle{plain}
\newtheorem{Theorem}{Theorem}

\newtheorem{Proposition}{Proposition}
\newtheorem{Corollary}{Corollary}

\newtheorem{Fact}{Fact}

\providecommand{\oo}[1]{\operatorname{o}\bigl(#1\bigr)}

\newcommand{\model}{EFRBM}
\newcommand{\IEEEQED}{$\square$}
\newcommand{\LMD}{\lambda_0,\dots,\lambda_n}

\makeatletter

{\theorembodyfont{\rmfamily} \newtheorem{Remark}{Remark}}
{\theorembodyfont{\rmfamily} }
{\theorembodyfont{\rmfamily} \newtheorem{Example}{Example}}
{\theorembodyfont{\rmfamily} }
{\theorembodyfont{\rmfamily} }

\newcommand {\R}{\mathbb R}

\newcommand{\be}{\begin{equation}}
\newcommand{\ee}{\end{equation}}

\newcommand{\Int}{\operatorname{{\mathrm Int}}}

\makeatletter
\renewcommand{\@biblabel}[1]{\quad#1.}
\makeatother

\date{}

\newcommand{\ben}{\begin{enumerate}}
\newcommand{\een}{\end{enumerate}}
\newcommand{\beqn}{\begin{align*}}
\newcommand{\eeqn}{\end{align*}}
\newcommand{\bi}{\begin{itemize}}
\newcommand{\ei}{\end{itemize}}

\usepackage{lastpage,fancyhdr,graphicx}

\date{}

\usepackage{lastpage,fancyhdr,graphicx}
\usepackage{epstopdf}
\begin{document}
\vspace*{0.2in}
\begin{flushleft}
{\Large
\textbf{A Deterministic Model for One-Dimensional Excluded Flow with Local~Interactions}}
\\ \bigskip
Yoram Zarai$^{1}$,
Michael Margaliot$^{2,\ast}$,
Anatoly B. Kolomeisky$^{3}$
\\
\bigskip
\bf{1} School of Electrical Engineering, Tel-Aviv
University, Tel-Aviv 69978, Israel.
\\
\bf{2}  School of Electrical Engineering and the Sagol School of Neuroscience, Tel-Aviv University, Tel-Aviv 69978, Israel.
\\
\bf{3} Department of Chemistry, Rice University, Houston, TX 77005-1892, USA.
\\
$\ast$ Corresponding Author. Email: michaelm@post.tau.ac.il
\end{flushleft}


\section*{Abstract}

Natural phenomena frequently involve a very large number of interacting molecules moving in confined regions of space. Cellular transport by motor proteins is an example of such collective behavior. We derive   a deterministic compartmental model for the unidirectional flow of  particles 
 along a one-dimensional lattice of sites with   nearest-neighbor interactions between the particles. The flow between consecutive sites is governed by a  ``soft'' simple exclusion principle  and by   attracting or repelling forces between  neighboring particles. Using tools from contraction theory, we prove that the model admits a unique steady-state and that every trajectory converges to this steady-state. 
Analysis and simulations of the effect of the attracting and repelling forces  on this steady-state
highlight the crucial  role that these forces may  play in   increasing the   steady-state flow,
and reveal  that this increase stems from the alleviation of traffic jams along the lattice. 
  Our theoretical analysis clarifies  microscopic aspects of complex multi-particle dynamic processes.



\section*{Introduction}\label{sec:intro}
Biological processes are governed by complex interactions between
 multiple particles that are confined in special compartments~\cite{Alberts2007}. 
One of the most important examples of such processes is   biological intracellular transport, which is carried by motor proteins (e.g., kinesins, dyneins, and myosins) \cite{kolomeisky_book}. These motor proteins, which are also known as biological molecular motors, can catalyze the reaction of adenosine triphosphate (ATP) hydrolysis, while at the same time converting the  energy produced during this chemical reaction into a mechanical work required for their movements along cellular filaments (such as microtubules and actin filaments)~\cite{kolomeisky_book}. 

Experimental observations clearly show that motor proteins usually function in large groups, suggesting that the interactions between the motors cannot be ignored~\cite{TASEP_book,kolomeisky2013motor}. Understanding  the
collective behavior of molecular motors is critical for uncovering mechanisms of complex  biological processes~\cite{kolomeisky_book,Chowdhury2005,team_motor}.  From a
theoretical point of view, intracellular transport processes are usually described using non-equilibrium multi-particle lattice models~\cite{TASEP_book}. In these models, the molecular motors are typically  represented by particles that hop along the lattice, and the lattice sites model
   the binding locations of  
	the motors along the filaments (or tracks). For a general  review on transport and traffic phenomena in biological systems see for 
example~\cite{kolomeisky_book,Chowdhury2005,TASEP_book,kolomeisky2013motor}.

A standard model from non-equilibrium statistical mechanics for molecular motors traffic (and numerous other processes) is the
\emph{totally asymmetric simple exclusion process}~(TASEP)~\cite{solvers_guide,TASEP_tutorial_2011,ped_survey}. 
This is also the standard model for ribosome flow during 
mRNA translation (see, e.g.~\cite{PhysRevLett.102.198104,Ciandrini2013,TASEP_tutorial_2011}). 
In~TASEP, particles  hop randomly   along a unidirectionally
 ordered lattice of sites. 
Simple exclusion means that a particle cannot move into a site that is already occupied by another particle, and thus each site can be either empty or occupied by a single particle. This models moving biological particles like ribosomes and motor proteins
 that have volume and thus cannot overtake a moving particle in front of them. 
 This   hard  exclusion principle  creates an intricate  indirect coupling between the particles.
In particular,  a slowly moving particle may lead to the formation of a traffic jam behind it.

To describe moving biological molecules with large sizes, a   version of  \emph{TASEP with extended objects} has been introduced and analyzed~\cite{Lakatos2003,Shaw2003,shaw2004local}. In this model, each particle covers~$\ell>1$ lattice
sites. Thus a particle
occupies  sites~$i,\dots,i+\ell-1$ for some~$i$,  and it  can hop to   site~$i+1$ provided that   site~$i+\ell$ is empty. This is used, for example, for modeling mRNA translation as it is known that every ribosome  (the particle) covers several  codons (sites) along the mRNA molecule~\cite{Shaw2003}.

There exist two 
     versions of~TASEP  that differ by their  boundary conditions. In~TASEP with \emph{open boundary conditions}
 the two sides of the chain are connected to two particle reservoirs with constant concentrations,
 and the particles can hop into the lattice chain (if the first site is empty) and out of the chain (if the last site is occupied). 
 In the open boundary \emph{homogeneous}   TASEP~(HTASEP), all the transition rates within the lattice are assumed to be equal and normalized to one, and thus the model is specified by an input rate $\alpha$, an exit rate $\beta$, and a parameter $N$ denoting the number of sites along  the lattice.
In TASEP with \emph{periodic boundary conditions} the chain is closed into a ring, and a particle that hops from the last site returns to the first site. TASEP   has been widely utilized for studying various natural and artificial  processes, including vehicular
 traffic flow, mRNA translation, surface growth, 
communication networks, and more~\cite{TASEP_book,tasep_ad_hoc_nets}.

 Ref.~\cite{turci2013transport} used HTASEP with periodic boundary conditions   to analyze transport on a lattice in the presence of local interactions between particles and substrate, illustrating the effect of local conformation of the substrate on the characteristics of the flow of molecular motors.
TASEP with particle interactions and with periodic boundary conditions was 
studied in~\cite{pinkoviezky2013modelling}, and with open boundary conditions in~\cite{antal2000asymmetric,hager2001minimal,teimouri2015theoretical,celis2015correlations}. Specifically, the authors in~\cite{teimouri2015theoretical,celis2015correlations} proposed a modified
 TASEP model that incorporates the realistic observed feature of nearest-neighbor interactions. In this model, the transition rate in every site along the lattice
 depends on the states of  four consecutive  sites. Their conclusions were
 that weak repulsive interaction results in maximal flux, and that the molecular motors are influenced more strongly by attractive interactions.  

Unfortunately, rigorous analysis of~TASEP is non-trivial, and exact solutions exist only in special cases, for example when considering the model with the \emph{homogeneous} rates~(HTASEP). Typically, the non-homogeneous 
case and cases that include other local interactions are only studied via various approximations and 
  extensive Monte Carlo computer simulations. These simulations are run until convergence to a (stochastic) steady-state, yet without a rigorous proof that convergence indeed takes place for all the feasible parameter values.

In this paper, we introduce a new \emph{deterministic}
 model for the flow of motor proteins along a one-dimensional lattice of sites with nearest-neighbor interactions between the motors. The flow of the motor proteins is unidirectional, and it satisfies  a ``soft'' simple exclusion principle. The nearest-neighbor effect is modeled by two ``force" interactions with parameters~$q$ and~$r$. It is more convenient to explain the effect of these interactions in   ``particle-like'' terms, although  in the new model the density in every site takes  values in the range~$[0,1]$ (and not~$\{0,1\}$). 

Consider a transition of a particle  from site~$i$   to site~$i+1$. If   site~$i+2$ is already occupied then the rate of movement depends on 
 a  parameter~$q\ge 0$  that represents an ``attachment/detachment force'' when generating \emph{new}  neighbors.  A  value~$q>1$ [$q<1$]
means that the particle will tend [not] to hop  forward, as there is a strong   attraction   [repulsion] to the particle in site~$i+2$.   On the other-hand, if   site~$i-1$ is already occupied then the rate of movement depends on
	a  parameter~$r\ge 0$  that represents an ``attachment/detachment force'' when breaking from \emph{old}  neighbors.  A    value~$r>1$ [$r<1$]
	means that the particle will tend [not] to hop forward, as there is a strong  repulsion   [attraction]     from the neighboring  particle in site~$i-1$. A value of~$q=1$ [$r=1$] implies no attachment/detachment force when generating new neighbors [when breaking from old neighbors].


An important advantage of our   model is that it is highly amenable to rigorous 
analysis even for \emph{non-homogenous} transition rates. We prove, for example,
that the dynamics  always converges to a steady-state density along the lattice. Thus, the flow also converges to a steady-state value.  
This steady-state depends on the lattice size, the transition rates,  and the parameters~$q$, $r$, but not on the initial
density along the lattice (i.e. the initial conditions). Analysis and simulations of the effect of the attracting and repelling forces 
 on this steady-state
highlight the crucial  role that these forces may  play in   increasing the   steady-state flow,
and reveal  that this increase stems from the alleviation of traffic jams along the lattice. 
It is well-known that  molecular motors indeed form traffic jams and that these have 
important biological implications (see, e.g.~\cite{PhysRevE.89.052703,Leduc17042012,kinesin_jams_disease}). 
In particular, analysis and simulations of the model reveal a new regime that may be interpreted as the ``opposite'' of a traffic jam along the lattice.

Our approach extends a deterministic
mathematical model that has been used
for describing and analyzing the flow of ribosomes along the mRNA molecule during the process of mRNA translation.
The next section provides a brief overview of this model. 

\subsection*{The Ribosome Flow Model (RFM)}
\begin{figure*}[t]
 \begin{center}
  \includegraphics[scale=0.8]{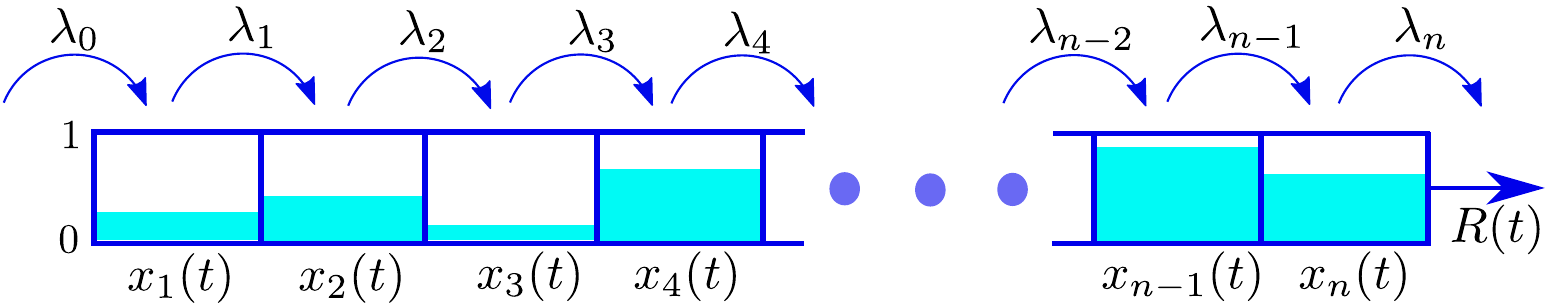}
\caption{The RFM models unidirectional flow along a chain of~$n$ sites.
The state variable~$x_i(t)\in[0,1]$ represents
 the density at site $i$ at time $t$. The parameter $\lambda_i>0$ controls the transition rate from  site~$i$ to site~$i+1$, with~$\lambda_0>0$ [$\lambda_n>0$] controlling    the initiation [exit] rate. The output rate at time $t$ is~$R(t) :=\lambda_n x_n(t)$. }\label{fig:rfm_diagram}
\end{center}
\end{figure*}

The  RFM~\cite{reuveni} is a nonlinear, continuous-time, compartmental model for the unidirectional flow of ``material" along a one-dimensional chain of $n$ consecutive compartments.  It can be derived via  a
 mean-field approximation of TASEP with open boundary conditions~\cite[Section 4.9.7]{TASEP_book}~\cite[p. R345]{solvers_guide}.
The~RFM includes $n+1$ parameters: $\lambda_0>0$ controls the initiation rate, $\lambda_n>0$ the exit rate, and $\lambda_i>0$, $i=1,\dots,n-1$, the transition rate from site~$i$ to site~$i+1$.
The state variable $x_i(t): \R_+ \to [0,1]$, $i=1,\dots,n$, describes 
the  normalized amount of ``material'' (or density) at site~$i$ at time~$t$, where~$x_i(t)=1$ [$x_i(t)=0$] indicates that site~$i$ is completely full [completely empty] at time~$t$.
Thus, the vector~$x(t):=\begin{bmatrix}x_1(t)&\dots&x_n(t)\end{bmatrix}'$ describes the density profile  along the chain at time~$t$.
The output rate at time $t$ is $R(t):=\lambda_n x_n(t)$~(see Fig.~\ref{fig:rfm_diagram}).

 Let $x_0(t)\equiv 1$, and $x_{n+1}(t)\equiv 0$. The dynamics of the RFM with $n$ sites is given by the following set of $n$ nonlinear ODEs:
 \be\label{eq:rfm_all}
 \dot{x}_i=\lambda_{i-1} x_{i-1}(1-x_i)-\lambda_i x_i (1-x_{i+1}), \quad i=1,\dots,n.
 \ee
This   can be explained as follows. The flow of material from site~$i$ to site~$i+1$ at time~$t$
is~$\lambda_{i} x_{i}(t)(1 - x_{i+1}(t) )$. This flow increases with the density at site~$i$, and decreases as site~$i+1$ becomes fuller. This corresponds to a ``soft''  version of a simple exclusion principle. Note that the maximal possible  flow  from site~$i$ to site~$i+1$  is the transition rate~$\lambda_i$. Thus Eq.~\eqref{eq:rfm_all} simply states that the change in the density at site $i$ at time $t$ is the input rate to site $i$ (from site $i-1$) at time $t$ minus the output rate (to site $i+1$) at time $t$.

The trajectories of the RFM evolve on the compact and convex state-space  
\[
      C^n:=\{x \in \R^n: x_i \in [0,1] , i=1,\dots,n\}.
\]
Let $\Int(C^n)$ [$\partial C^n$] denote the interior [boundary] of $C^n$.
Ref.~\cite{RFM_stability} has shown that the RFM is a
\emph{tridiagonal cooperative dynamical system}~\cite{hlsmith},
and consequently~\eqref{eq:rfm_all}
admits a \emph{unique} steady-state density~$e=e(\LMD) \in \Int(C^n)$ that is globally asymptotically stable, that is, $\lim_{t\to\infty} x(t,a)=e$  for all $a\in C^n$ (see also~\cite{RFM_entrain}). This means that trajectories corresponding to different initial conditions all converge to the same steady-state density~$e$. 
In particular, the density at the last site~$x_n(t)$ converges to the value~$e_n$, so the output rate~$R(t)$ converges to a steady-state  value~$R:=\lambda_n  {e}_n$. 

An important advantage of the RFM (e.g. as compared to TASEP) is that it is amenable to mathematical analysis using tools from systems and control theory. Furthermore, most of the analysis hold for the general, non-homogeneous case (i.e. the case  where the transition rates~$\lambda_i$ 
   differ from one another). For more on the analysis of the RFM and its  biological implications, 
	see~\cite{HRFM_steady_state,zarai_infi,RFM_stability, RFM_feedback, RFM_entrain, RFM_concave,RFM_sense, RFMR,rfm_control,rfm_opt_down,RFM_r_max_density}.  

In this paper, we extend
 the RFM to include nearest-neighbor interactions, namely, 
   binding and repelling actions that are dynamically activated for each site
	based on the state of its neighboring sites. 
	A parameter $r$ [$q$] controls
	the binding/repelling  forces between   two existing
		[new] neighbors. We refer to the new model
	as the \emph{excluded flow with local repelling and binding model}~({\model}). It is important to note that this is significantly 
	different from the~RFM. For example, the
	{\model}, unlike the RFM, is \emph{not}
	a cooperative system~\cite{hlsmith}.
	Also, in the RFM the dynamics at site $i$ is directly
	affected by its two  nearest neighbors sites, whereas in the {\model} 
	the dynamics is directly affected by the density in four neighboring 
	sites. Thus, unlike the RFM, the {\model} is not a tridiagonal system.
	Also, the RFM has been used to model ribosome flow, whereas here
	we apply the {\model} to study the flow of motor proteins.

We show that the {\model} is a  contractive dynamical system. This holds for any set of feasible transition rates and local interaction 
forces including the case of non-homogeneous transition rates. 
This implies that the {\model} admits a unique steady-state
   that is globally asymptotically stable. Thus, every set of parameters corresponds to a unique steady-state 
	output rate. 
	We analyze the behavior of this steady-state under the assumption~$rq=1$  
	that follows from  fundamental thermodynamic arguments  (see~\cite{anataoly_inter_theor}).
	We show that a small neighbor-repelling force (i.e. small~$r$ and thus a large~$q=1/r$)
	leads to a   small output rate. Analysis and simulations show that this is due to the formation of traffic jams at the beginning of the lattice.
	On the other-hand, a strong neighbor-repelling force (i.e. large~$r$ and small~$q$) lead to a high output rate. 
	In this case, an interesting phenomena emerges: the density in every second site goes to zero. 
	This ``separation of densities'' is the ``opposite'' of a traffic jam. These results highlight the impact
	of traffic jams
	on  the output rate. 
	 
The remainder of this paper is organized as follows. The next section describes
 the~{\model}. The following two sections describe our main analysis results and their biological implications.
This includes analysis of the  asymptotic behavior of the~{\model}, and
 the effects of the nearest-neighbor interactions on the steady-state behavior of the~{\model}. 
The final section summarizes and describes several directions for further research.
To increase the readability of this paper, all the proofs are placed in the Appendix.

\section*{The {\model}}
The {\model} with  $n$ sites includes~$n+3$   parameters:
\begin{itemize}
\item $\lambda_i>0$, $i=0,\dots,n$, controls the transition rate from site $i$ to site $i+1$, where $\lambda_0$ [$\lambda_n$] controls the input [output] rate.
\item $r\ge0$ is the attachment/detachment  force between any two existing (consecutive) neighbors. 
\item $q\ge0$ is the attachment/detachment  force between any two new (consecutive) neighbors.
\end{itemize}

\begin{figure*}[t]
\centering
 \includegraphics[scale=0.75]{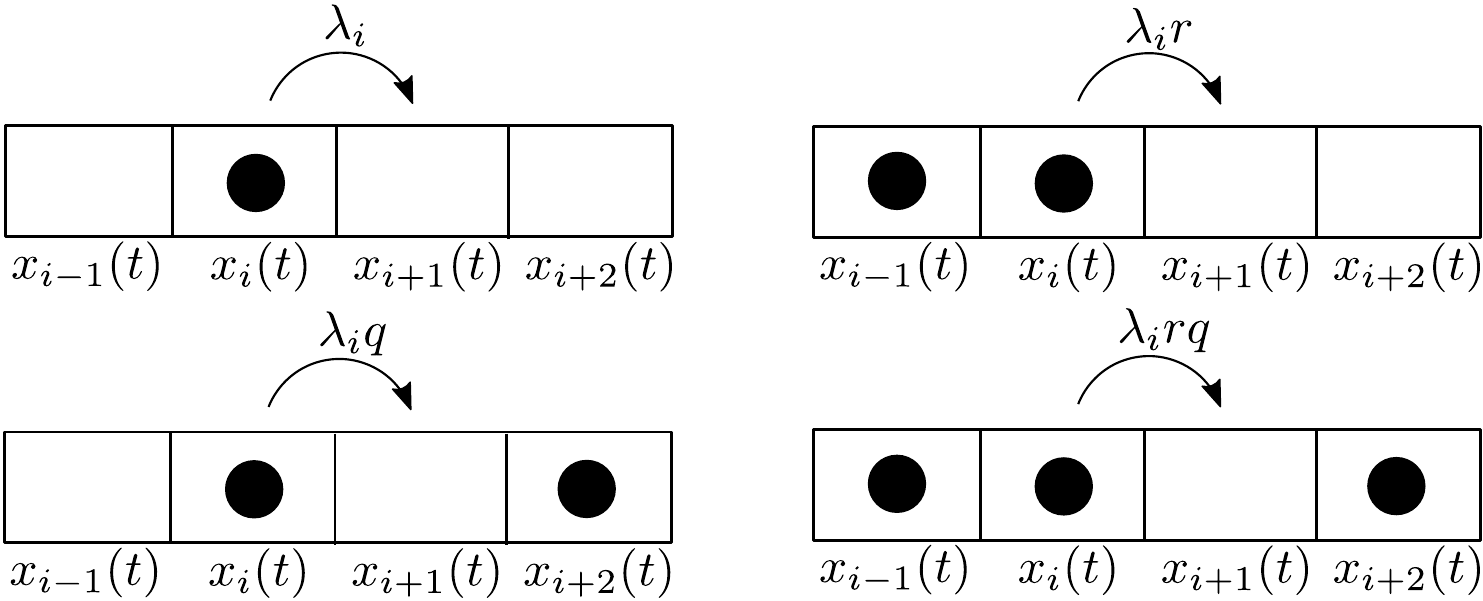}
\caption{A schematic explanation of
the transition flow from site~$i$ to site~$i+1$ in the {\model}. Upper-left: when both sites~$i-1$ and~$i+2$ do not contain particles, the transition rate is~$\lambda_i$. Lower-left: when site~$i-1$ does not contain particles, and site~$i+2$ does, the transition rate is~$\lambda_i q$. Upper-right: when site $i-1$ contains particles, and site $i+2$ does not, the transition rate is $\lambda_i r$. Lower-right: when both sites $i-1$ and $i+2$ contain particles, the transition rate is $\lambda_i r q$.  }\label{fig:rfmi_sitei}
\end{figure*}

Fig.~\ref{fig:rfmi_sitei} depicts the four possible transition   scenarios from site~$i$ to 
site~$i+1$, and the rates in each case.
For simplicity, we  use a schematic ``particle-like'' explanation, although in the~{\model} the state-variables represent a normalized
 material density in the range~$[0,1]$ and not a binary choice~$\{0,1\}$  like in~TASEP. 
 If both sites $i-1$ and $i+2$ do not contain particles, the transition rate is simply~$\lambda_i$, as in the RFM.  If a particle  is located
   at site $i-1$ [$i+2$] but site~$i+2$ [$i-1$] is empty then the transition rate is~$\lambda_i r$ [$\lambda_i q$].
If both sites   contain particles the transition rate is~$\lambda_i r q$.

The {\model} also includes~$n$ state-variables  $x_i(t)$, $i=1,\dots,n$. Just like in the RFM,
$x_i(t)$ 
  describes the normalized density at site~$i$ at time~$t$, 
where~$x_i(t)=0$ [$x_i(t)=1$] means that the site is completely empty [full]. 

To state the dynamical equations describing the {\model}
we introduce more notation.
Let $x_0(t)\equiv1$, $x_{n+1}(t)\equiv0$, and denote
\be\label{eq:defz}
									z_i(t):=\begin{cases}  x_i(t), & i=1,\dots,n,\\
												                  0, & \text{otherwise}. \end{cases}
\ee
Then the {\model} is described by 
\be\label{eq:rfmi_odes}
\dot x_i = g_{i-1}(x)-g_i(x), \quad i=1,\dots,n,
\ee
where
\be\label{eq:rfmi_g}
g_i(x) := \lambda_i x_i(1-x_{i+1})  (1+(q-1)z_{i+2})(1+(r-1)z_{i-1} ) .
\ee
We now explain these equations. The term~$g_i(x)$ represents the flow from site~$i$ to site~$i+1$,
so Eq.~\eqref{eq:rfmi_odes} means that the change in the density at site~$i$ 
is the   inflow from site $i-1$ minus the outflow  to site~$i+1$. 
To explain Eq.~\eqref{eq:rfmi_g}, consider for example the case~$i=2$ (and assume that~$n\geq 4$). Then~\eqref{eq:rfmi_g} yields
\be\label{eq:g2xp}
g_2(x)  = \lambda_2 x_2(1-x_{3})  (1+(q-1)x_{4})(1+(r-1)x_{1} ) .
\ee
The term~$x_2$ means that the flow from site~$2$ to site~$3$ increases with the density at site~$2$. 
The term~$(1-x_3)$ represents soft exclusion: as the density at site~$3$ increases, the transition from site~$2$ to site~$3$ gradually
decreases. 
The term~$(1+(q-1)x_{4})$ represents the fact that the flow into site~$3$ also depends on the density at site~$4$: 
if~$q>1$ [$q<1$] then the transition increases [decreases] with~$x_4$, that is, the ``particles'' at site~$4$ ``attract'' [``repel'']
 the particles that move  from site~$2$ to site~$3$. The term~$(1+(r-1)x_{1} )$ is similar but  represents an
attachment/detachment force between the ``particles'' in sites~$1$ and~$2$.

Note that  for $r=q=1$, $g_i(x)=\lambda_i x_i(1-x_{i+1})$, and thus in this case the {\model} reduces to 
 the~RFM (see~\eqref{eq:rfm_all}).
On the other hand, if $q=r=0$ 
then~$g_i(x)=\lambda_i x_i(1-x_{i+1})  (1-x_{i+2})(1-x_{i-1})$. This represents a kind of  an ``extended objects''
 RFM, as the transition from site~$i$ to site~$i+1$ decreases with the density in sites~$i-1$, $i+1$, and~$i+2$.

\begin{Remark}\label{rem:tvrates}
It is useful to think of the {\model} as an RFM with \emph{time-varying} transition rates. For example, we can write
\eqref{eq:g2xp} as
\[
g_2(x(t))  = \eta_2(t) x_2(t)(1-x_{3}(t))  ,
\]
where~$ \eta_2(t):= \lambda_2    (1+(q-1)x_{4}(t))(1+(r-1)x_{1}(t) ) $. 
Note that this time-varying transition rate depends on~$\lambda_2$ (i.e.,  the fixed site to site transition rate), and 
also on $r$ and $q$ and    the time-varying densities  in the neighboring sites, as these determine  the interaction
 forces between the moving particles. 
\end{Remark}

We denote  the flow from site~$x_n$ to the environment  by 
\begin{align}\label{eq:rfmi_Rt}
R(t)&:= 
 \lambda_n x_n(t) (1+(r-1)x_{n-1}(t)).
\end{align}
This is the  \emph{output rate} at time~$t$.

\begin{Example}
The  {\model} with~$n=3$ sites is given by: 
\begin{align}\label{eq:3dim}
										\dot x_1 & = \lambda_0(1-x_1)(1+(q-1)x_2)
										-\lambda_1 x_1 (1-x_2)(1+(q-1) x_3) , \nonumber \\
										\dot x_2 & = \lambda_1x_1(1-x_2)(1+(q-1)x_3)  
											-\lambda_2 x_2 (1-x_3) (1+(r-1)x_1)    , \nonumber \\
\dot x_3 & =  \lambda_{2} x_{2}   (1-x_3)    (1+(r-1)x_1)  
    -\lambda_3 x_3 (1+(r-1)x_{2})    . 
\end{align}
If~$q=r=0$ then this becomes
\begin{align}\label{eq:3dim_zero}
										\dot x_1 & = \lambda_0(1-x_1)(1-x_2)   -\lambda_1x_1(1-x_2)(1-x_3)  , \nonumber \\
										\dot x_2 & = \lambda_1x_1(1-x_2)(1-x_3)      -\lambda_2(1-x_1)x_2(1-x_3)   
										    , \nonumber \\
\dot x_3 & =  \lambda_{2}(1-x_{1})x_{2}(1-x_3)    
    -\lambda_3(1-x_{2})x_3  
										  . 
\end{align}
On the other-hand, for   $q=1$ and~$r=0$ \eqref{eq:3dim} becomes
\begin{align}\label{eq:3dimrqspec}
	\dot x_1 & = \lambda_0(1-x_1)  -\lambda_1x_1(1-x_2) , \nonumber \\
										\dot x_2 & = \lambda_1x_1(1-x_2)     -\lambda_2(1-x_1)x_2(1-x_3)   
										   , \nonumber \\
\dot x_3 & =  \lambda_{2}(1-x_{1})x_{2}(1-x_3)     
    -\lambda_3(1-x_{2})x_3 , 
\end{align}
and this system admits a continuum of  \updt{steady-states}, as~$\begin{bmatrix}  1& 1& s \end{bmatrix}'$ is a \updt{steady-state} for all~$s$.~\hfill{$\square$} 
\end{Example}

\updt{Following~\cite{anataoly_inter_theor} (see also~\cite{nadler_schulten1986}), we view
creating and breaking a pair of particles 
 as opposite chemical transitions, so by detailed balance arguments:
$\frac{q}{r}=\exp\left(\frac{E}{K_BT}\right)$,
where~$E$ is the interaction energy. As in~\cite{anataoly_inter_theor}, we also assume that~$E$ is equally split between the creation
and breaking processes, so
\be\label{eq:qrpropd}
q=\exp\left(\frac{E}{2K_BT}\right),\quad r=\exp\left(\frac{-E}{2K_BT}\right).
\ee
This has a clear physical meaning. If~$E>0$ the   interaction is attractive,  so the
particle moves faster when creating a new pair~$(q>1)$ since the energy of the system decreases by~$E$. On the other-hand, breaking out of the cluster increases the energy by~$E$ and the transition rate is thus slowed down~($r<1$). 
Similarly, the case~$E<0$ corresponds to a repulsive interaction and then~$q<1$ and~$r>1$. }
Note that~\eqref{eq:qrpropd} implies in particular that
\be\label{eq:rq1}
rq=1.
\ee
In this case, the~{\model} contains $n+2$ parameters: $\LMD$, and $r$ (as $q =1/r$).
Note that if~\eqref{eq:rq1} holds then~\eqref{eq:rfmi_g} becomes
\be\label{eq:rfmi_g_rq1}
g_i(x) = \lambda_i x_i(1-x_{i+1})  (1-\frac{r-1}{r}  z_{i+2})(1+(r-1)z_{i-1}) .
\ee

The next section   derives several theoretical results on the dynamical
behavior of the~{\model}. 
Recall that all the proofs are placed in the Appendix.

\section*{Asymptotic behavior of the~{\model}}\label{sec:asy}
 Let~$x(t,a)$ denote the solution of the~{\model}
at time~$t$ for the initial condition~$x(0)=a\in C^n$.

\subsection*{Invariance and persistence}
The next result shows that the $n$-dimensional 
  unit cube~$C^n$ is an invariant set of the~{\model}, that is, any trajectory that emanates
	from
	an initial condition in~$C^n$ remains in~$C^n$ for all time. 
	Furthermore,
any trajectory emanating from the boundary of~$C^n$ ``immediately  enters''~$C^n$. 
This is a technical result, but it is important as in the interior of~$C^n$ the 
{\model} admits several useful properties. 
\begin{Proposition}\label{prop:inv}
Assume that~$q,r >0$. 
For any~$\tau>0$ 
there exists $d=d(\tau)\in(0,1/2)$ such that 
\[
d\le x_i(t+\tau,a)\le 1-d,
\]
for all $a\in C^n$, all $i\in\{1,\dots,n\}$, and all $t\ge 0$.
\end{Proposition}
This means that all  the  trajectories of the {\model} enter and remain in the interior of~$C^n$ 
after an arbitrarily  short time.  In particular, 
 both $C^n$ and $\Int(C^n)$ are invariant sets of the {\model} dynamics.

From a biological point of view  this means that if the system is initiated such that every density is in~$[0,1]$
 then this remains true for all time~$t\geq 0$, so the equations ``make sense'' in this respect. Furthermore,
after an arbitrarily short time the densities  are all in~$(0,1)$, i.e. any completely empty [full] site 
immediately becomes not  completely empty [full].



\subsection*{Contraction}
Differential analysis and in particular contraction theory  proved to be a powerful tool for analyzing the asymptotic
behavior
of nonlinear dynamical systems. In a contractive system, trajectories that emanate from different initial conditions approach
 each other at an exponential rate~\cite{LOHMILLER1998683,entrain2011,sontag_contraction_tutorial}.

For our purposes, we require a   generalization of 
contraction with respect to (w.r.t.) a fixed norm that has been
introduced in~\cite{3gen_cont_automatica}. 
Consider the time-varying dynamical system:
\be\label{eq:tvsys}
\dot{x}(t)=f(t,x(t)),
\ee
whose trajectories evolve on an invariant set~$\Omega\subset\R^n$ that is compact and convex.  Let~$x(t,t_0,a)$ denote the
solution of~\eqref{eq:tvsys} at time~$t$ for the initial condition~$x(t_0)=a$.
 The dynamical system~\eqref{eq:tvsys}
is said to be   \emph{contractive after a small  overshoot}~(SO)~\cite{3gen_cont_automatica}
on~$\Omega$  w.r.t. a norm~$|\cdot|:\R^n\to\R_+$ if for any~$\varepsilon>0$
there exists~$\ell=\ell(\varepsilon)>0$
 such that
\[
|x(t ,t_0,a)-x(t,t_0 ,b)|\leq(1+\varepsilon)  \exp(-  (t-t_0) \ell)|a-b|,
\]
for all $a,b\in\Omega$ and all~$t\geq t_0\geq 0$.
Intuitively speaking, this means that any two trajectories of the system
approach  each other at   an exponential rate~$\ell$,  but with  an arbitrarily
 small  overshoot of~$1+\varepsilon$.

Let~$|\cdot|_1:\R^n\to\R_+$ denote the~$L_1$ norm, i.e. for~$z\in\R^n$,~$|z|_1=|z_1|+\dots+|z_n|$.
 \begin{Proposition}\label{prop:weak_cont}
The {\model} with $q,r>0$ is SO on~$C^n$ w.r.t. the~$L_1$ norm, that is, for any~$\varepsilon>0$ there
	exists~$\ell=\ell(\varepsilon)>0$ such that
	\be\label{eq:sorfm}
|x(t ,a)-x(t ,b)|_1\leq(1+\varepsilon)  \exp(-\ell t)|a-b|_1,
\ee
for all $a,b\in C^n$ and all~$t\geq0$.
\end{Proposition}

From a biological point of view this means the following. 
The state of the system at any time~$t$ is a vector describing 
 the density at each site at time~$t$.  
We measure the distance between any  two density vectors using the~$L_1$ vector norm. 
Suppose that we initiate the system with two different densities.  This generates two different
solutions of the dynamical  system. The distance between these solutions  decreases with time at an exponential rate.

The next example demonstrates this 
 contraction property. Let~$1_n$ [$0_n$]
denote the column vector of~$n$ ones [zeros]. 
 
\begin{figure}[t]
  \begin{center}
  \includegraphics[scale=0.6]{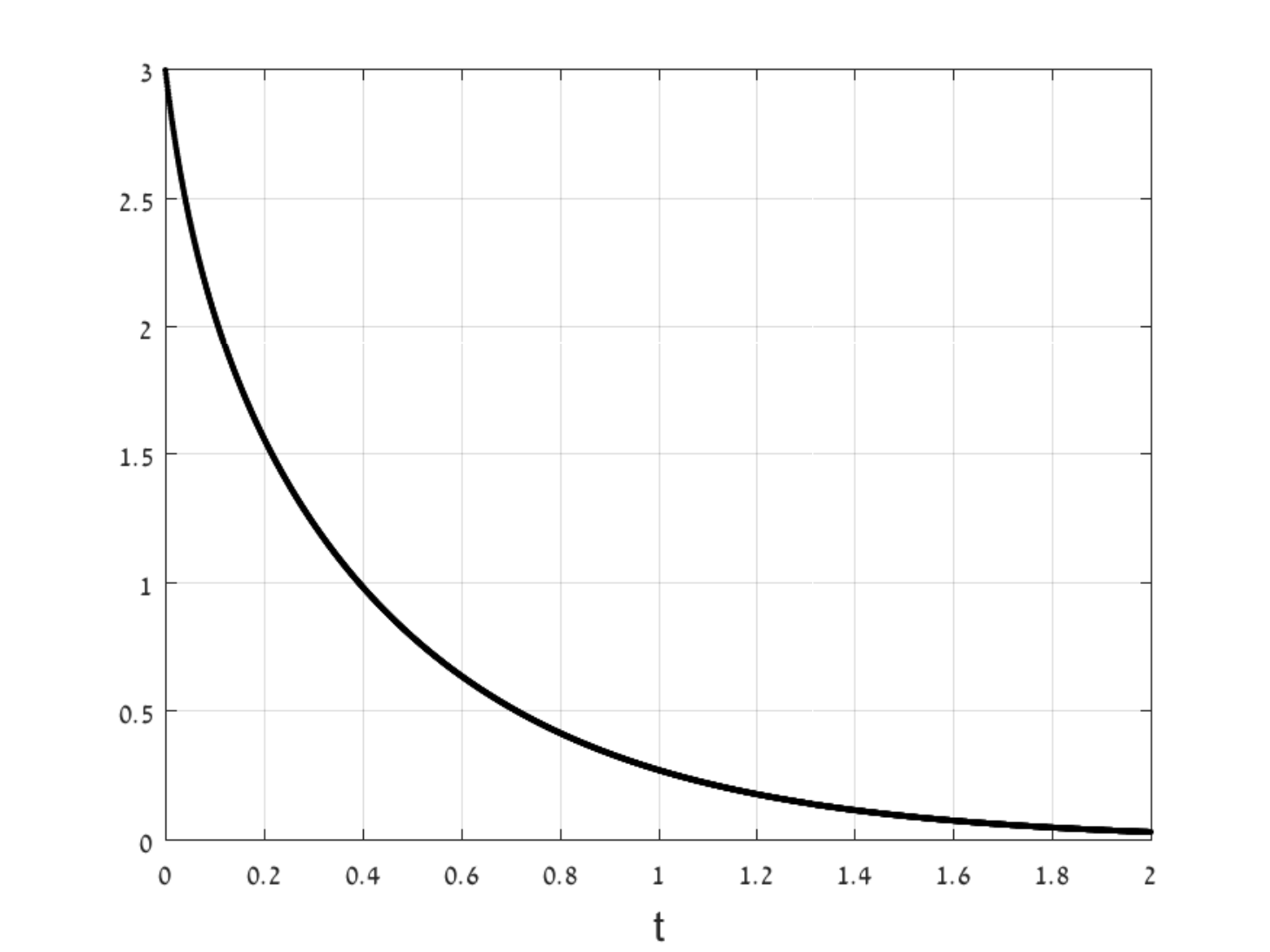}
  \caption{The distance~$|x(t,a)-x(t,b)|_1$ as a function of time for the {\model} in Example~\ref{exp:dist_e}.}\label{fig:rfmi_n6_dist_e}
  \end{center}
\end{figure}

\begin{Example}\label{exp:dist_e}
Consider the {\model} with dimension~$n=3$, and parameters~$\lambda_0=1$, $\lambda_1=2$, $\lambda_2=3$, $\lambda_3=4$,~$r =5$, and~$q=1/5$. 
Fig.~\ref{fig:rfmi_n6_dist_e} depicts $|x(t,a)-x(t,b)|_1$, with $a =0_3$ and~$b=1_3$,
 as a function of time for $t\in[0,2]$.
It may be seen that the $L_1$ distance between the two trajectories  goes   to zero at an exponential rate.~\hfill{$\square$}
\end{Example}

Prop.~\ref{prop:weak_cont}  implies  that the {\model} satisfies  several important asymptotic properties. These are described in the following subsections.

\subsection*{Global asymptotic stability}
Write the   {\model}~\eqref{eq:rfmi_odes}  as $\dot{x}=f(x)$. Since the compact and convex set~$C^n$ is an invariant set of the dynamics, it contains at least one  steady-state. That is, there exists~$e=e(\LMD, q,r)$ such that~$f(e)=0_n$.  
By Proposition~\ref{prop:inv},~$e\in\Int(C^n)$. 
Using~\eqref{eq:sorfm} with $b:=e$ yields the following result.
\begin{Corollary}\label{cor:asymp}
Assume that~$q,r>0$. Then the   {\model} admits a unique \updt{steady-state}~$e\in\Int(C^n)$ that is globally asymptotically stable, i.e. 
\[
\lim_{t\to\infty} x(t,a)=e, \text{ for all } a \in C^n.
\]
\end{Corollary}

This means that   any
   solution of the~{\model}   converges to a  unique steady-state density (and thus a unique steady-state output rate)  that
	depends on the rates~$\lambda_i$, and the parameters~$r$ and $q$, but not on the initial condition. From a biological point of view, this means that the system
	always converges to a steady-state density and  a corresponding
	steady-state output rate,
	and thus 
	it makes sense to study how these depend  on the various parameters. 
	
	Note that the assumption that~$r,q>0$ cannot be dropped. Indeed, Eq.~\eqref{eq:3dimrqspec}, corresponding to a {\model} 
	with~$n=3$, $q=1$ and~$r=0$,  admits a continuum
	of
\updt{steady-states}. 

\begin{figure}[t]
  \begin{center}
  \includegraphics[scale=0.4]{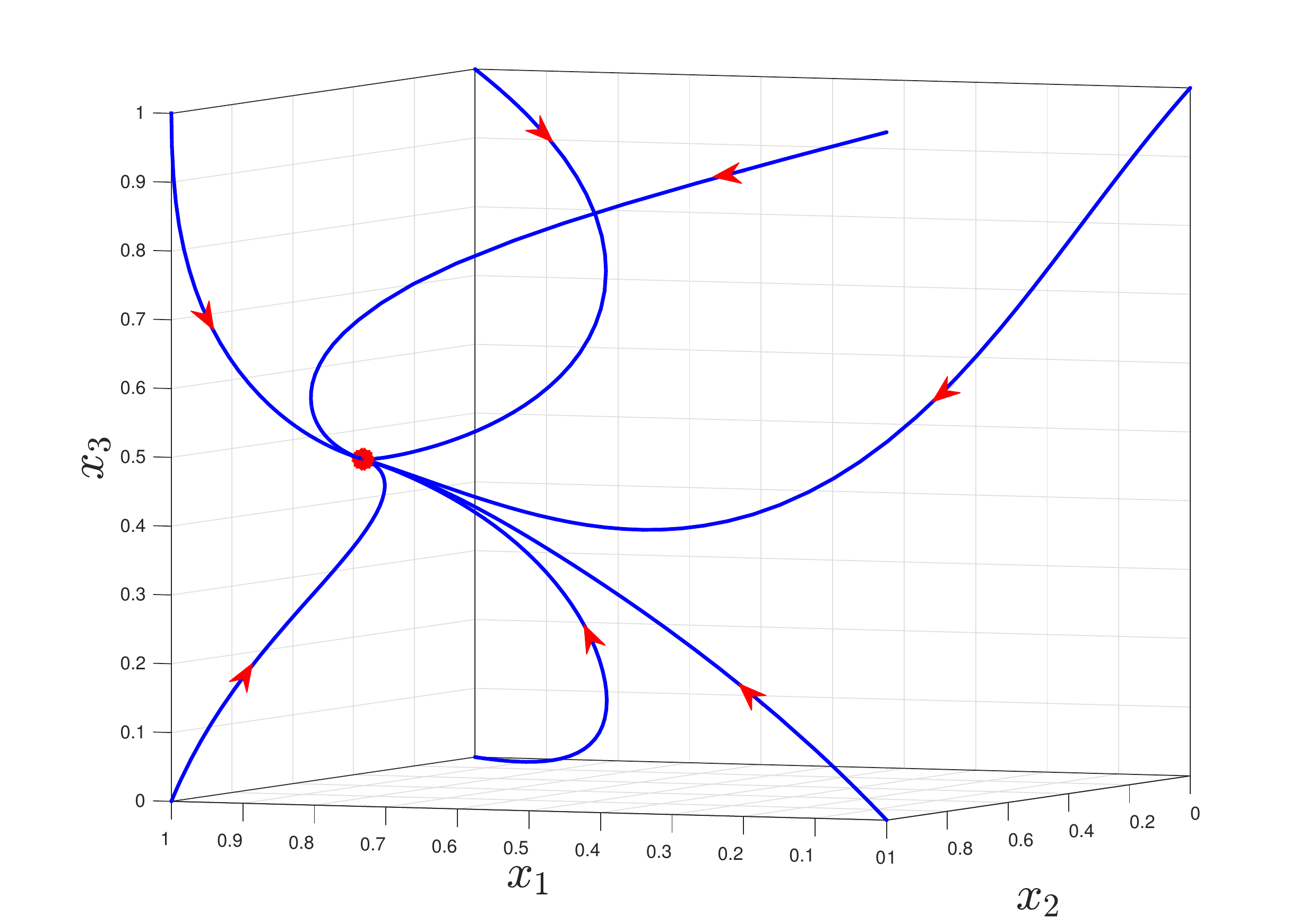}
  \caption{Trajectories of the {\model} in Example~\ref{exp:exp1} for seven arbitrary initial conditions. 
	The \updt{steady-state}~$e$ is denoted by~$*$. }\label{fig:rfmi_n3_ss}
  \end{center}
\end{figure}

\begin{Example}\label{exp:exp1}
Fig.~\ref{fig:rfmi_n3_ss} depicts the trajectories of~\eqref{eq:rfmi_odes} with~$n=3$, $\lambda_0=0.5$, $\lambda_1=0.8$, $\lambda_2=0.7$, $\lambda_3=0.6$,~$r =1/2$, and~$q=2$, for several initial conditions.
 It may be seen that all trajectories
 converge to a unique \updt{steady-state}~$e =\begin{bmatrix} 0.8555   & 0.7881 &    0.4268\end{bmatrix}'$.
(All the numerical values in the simulations described in  this paper are to four digit accuracy.)~\hfill{$\square$}
\end{Example}

The rigorous proof that every trajectory converges to a steady-state is important, as it implies that
after some time the densities are very close to their steady-state values. 
The next step is to analyze this steady-state density and the corresponding steady-state output 
 rate, and explore how these are related to the various parameters of the model. 

\subsection*{Analysis of the steady-state}
At steady-state, (i.e. for $x=e$) the left-hand side of all the equations in~\eqref{eq:rfmi_odes} is zero (i.e. $\dot{x}_i=0$, $i=1,\dots,n$), so
$g_{i-1}(e)=g_i(e)$ for all~$i$. This implies that 
\begin{align}\label{eq:eq_of_model}
										    \lambda_0 & (1-e_1)  (1+(q-1)e_2)         \nonumber  \\
										& = \lambda_1 e_1(1-e_2)    ( 1+(q-1)e_3)       \nonumber \\
										  & =  \lambda_2  e_2 (1-e_3)(1+(q-1)e_4)(1+(r-1)e_1) \nonumber \\
         & =    \lambda_3   e_3(1-e_4)(1+(q-1)e_5)(1+(r-1)e_2)         \nonumber \\
										&\vdots   \\
  & =  \lambda_{n-1}  e_{n-1} (1-e_n)             (1+(r-1) e_{n-2})        \nonumber \\
   &=  \lambda_n    e_n   (1 +(r-1)e_{n-1})       ,\nonumber 
\end{align}
and also that the steady-state flow satisfies 
\be\label{eq:rfmi_R_eq}
R = g_i(e), \quad i=0,\dots,n . 
\ee
In particular,~$R= \lambda_n    e_n   (1 +(r-1)e_{n-1})$  and since every~$e_i \in (0,1)$,
the steady-state flow is positive (i.e. a left-to-right flow) for any~$r>0$.

Also, for the case~$r q=1$ it follows from~$R= \lambda_0   (1-e_1)  (1+(q-1)e_2)$ that for~$r\geq 1$, $R \leq \lambda_0$,
whereas for~$r< 1$ it follows from~$R= \lambda_n    e_n   (1 +(r-1)e_{n-1}) $
 that~$R \leq \lambda_n$, so 
\[
				R\leq \max\{\lambda_0 ,  \lambda_n \}.
\]  
This means in particular that the output rate is always bounded.

\begin{Fact}\label{fct:homo}
It follows from~\eqref{eq:eq_of_model} that if we multiply all the $\lambda_i$s by a parameter $c > 0$ then $e$ will not change, i.e. $e(c\lambda)=e(\lambda)$. Thus, by~\eqref{eq:rfmi_R_eq} $R(c\lambda) = cR(\lambda)$, for all $c > 0$, that is, the steady-state flow [density] is \emph{homogeneous of degree one [zero]} w.r.t. the $\lambda_i$s.
\end{Fact}

In the spacial case~$n=2$   the steady-state equations~\eqref{eq:eq_of_model} can be solved in closed-form.
\begin{Fact}\label{fact:case2simp}
Consider the~{\model} with~$n=2$ and~$q=1/r$. 
Define
\begin{align}\label{eq:defais}
a_1&:=  (1-\frac{1}{r})(\lambda_2 r+\lambda_1)  +\frac{\lambda_1 \lambda_2}{\lambda_0}.
\end{align}
Then~$e=\begin{bmatrix} e_1& e_2 \end{bmatrix}' $ 
is given by
\begin{align}\label{eq:expa2}
e_2&=\frac{  \lambda_1+\lambda_2+a_1- \sqrt{ (\lambda_1+\lambda_2+a_1)^2-4 a_1 \lambda_1  }   }{2 a_1},\nonumber \\
	e_1&=\frac{\lambda_2 e_2}{\lambda_1+( \lambda_2(1-r)-\lambda_1  )e_2 }.
\end{align}
Note that even in this case the expression for~$e$ is non-trivial. 
\end{Fact}

Let~$\R^n_{++}$ denote the set of~$n$ dimensional vectors with all entries positive. 
 Let~$v:=\begin{bmatrix}
\lambda_0& \dots& \lambda_n & r & q \end{bmatrix}'$ denote the set of parameters in the~{\model} with dimension~$n$.
The results above imply that there exists a function~$h:\R^{n+3}_{++} \to \Int(C^n)$ 
such that~$e=h(v)$ is the unique steady-state of the {\model} with
  parameters~$v$.  
\begin{Proposition}\label{prop:ediff}
The function~$h:\R^{n+3}_{++} \to \Int(C^n)$ is analytic. 
\end{Proposition}
This result allows in particular   to consider 
 the derivatives of the steady-state density~$e=e(v)$ and the steady-state output rate~$R=R(v)$
 w.r.t. small changes in some of the parameters~$v$, that is, the sensitivity of the steady-state w.r.t. small changes in the parameters.

\section*{Effect of nearest-neighbor interactions }\label{sec:rates}

\begin{figure}[t]
  \begin{center}
  \includegraphics[width= 8cm,height=7cm]{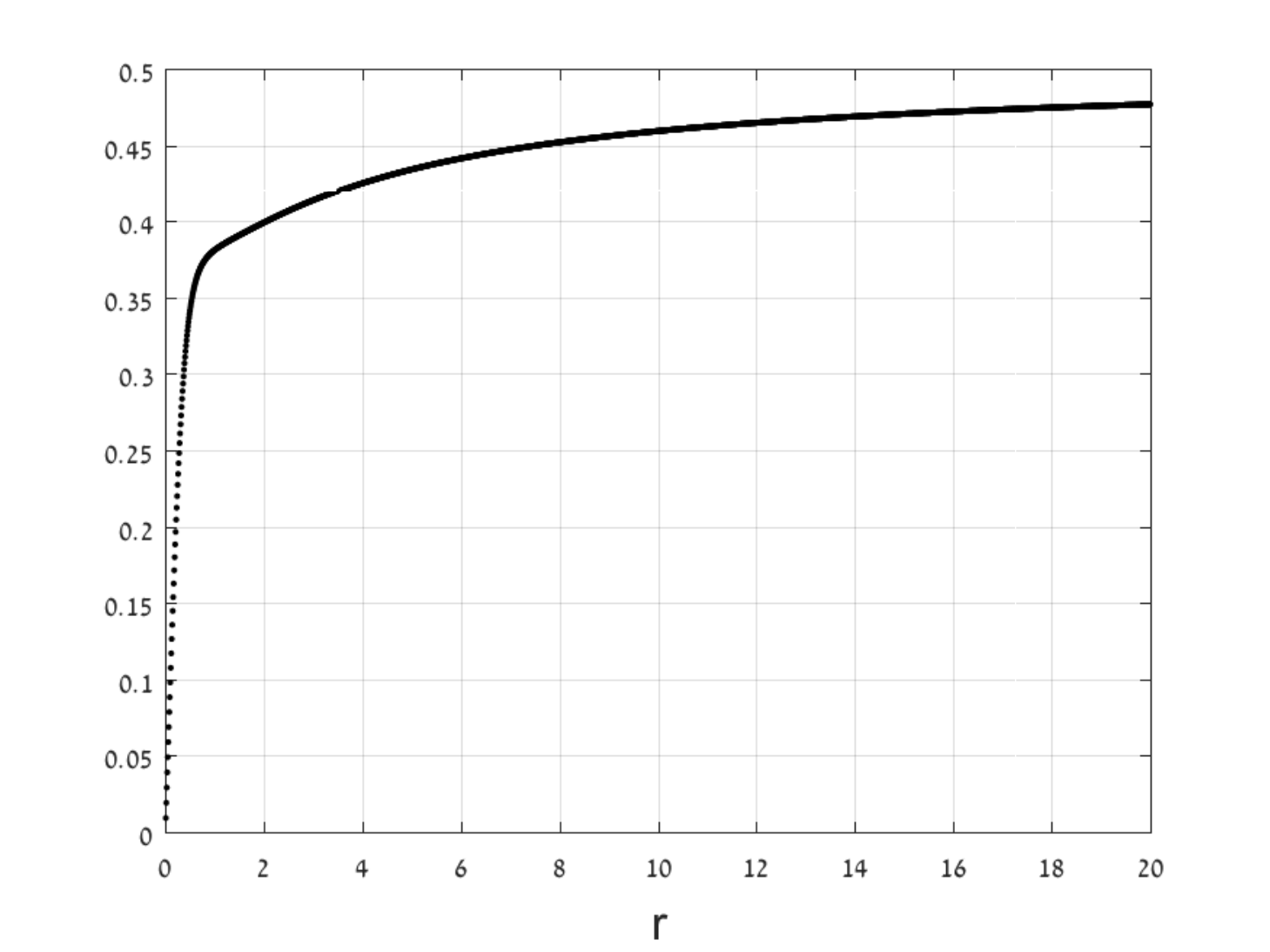}
 \caption{Steady-state output rate~$R$
 as a function of~$r\in[0.01,20]$ for a {\model} with~$n=2$,~$\lambda_i=1$ for all~$i$, and~$q=1/r$. Note that the value for~$r=1$ is the steady-state output rate in the~RFM. }\label{fig:floqrq}
  \end{center}
\end{figure}

We begin with several simulations demonstrating the 
effect of the parameter~$r$ (and~$q=1/r$) on the   steady-state of the~{\model}. 
\begin{Example}
   Consider  a~{\model} with~$n=2$ and rates~$\lambda_0=\lambda_1=\lambda_2=1$.  
 Fig.~\ref{fig:floqrq} depicts the steady-state
 output rate~$R$   
as a function of~$r$.
 It may be seen that~$R$      monotonically increases with~$r$.
In particular, for~$r=1$ (i.e., the RFM) $R=0.3820$, wheres for~$r=20$, $R=0.4778$,
 that is, the steady-state flow is increased by about~$25\%$. 
 When considering the comparison with the~RFM, one should bear in mind that the~{\model} corresponds to an RFM
with time-varying rates~$\eta_i(t)$  that may effectively be much higher than the fixed rates~$\lambda_i$. 
We assume that the energy that is needed to generate these higher rates comes from the additional 
interaction forces between the particles.~\hfill{$\square$}
 \end{Example}

\begin{figure}[t]
  \begin{center}
  \includegraphics[width= 8cm,height=7cm]{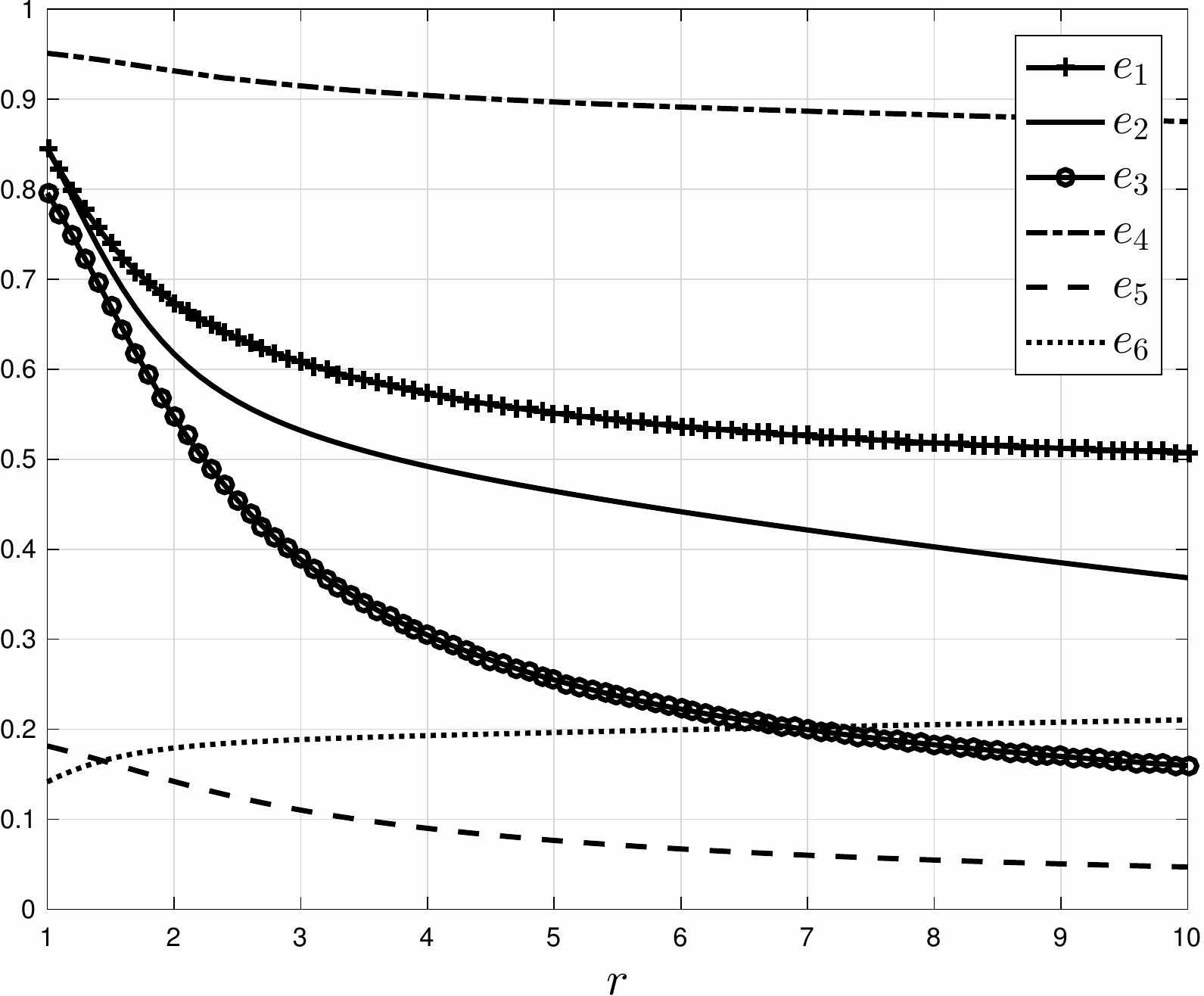}
 \caption{Steady-state densities~$e_i$ as a function of~$r\in[1,10]$ for a~{\model} with~$n=6$,~$\lambda_0=1.0$, $\lambda_1=1.2$, $\lambda_2=0.9$, $\lambda_3=4.0$, $\lambda_4=0.2$, $\lambda_5=1.0$, $\lambda_6=1.1$, and~$q=1/r$. Note that as~$r$ increases all densities become much smaller than one, that is, there are no traffic jams.}\label{fig:rfmi_n6_e}
  \end{center}
\end{figure}

The next example demonstrates that the  increase in~$R$ as~$r$ increases is 
because the  neighbor-repelling forces lead 
to an 
alleviation 
of  traffic jams.
\begin{Example}
Consider the {\model} with dimension $n=6$, $\lambda_0=1.0$, $\lambda_1=1.2$, $\lambda_2=0.9$, $\lambda_3=4.0$, $\lambda_4=0.2$, $\lambda_5=1.0$, and $\lambda_6=1.1$. Consider first the case $r=q=1$ (i.e., the RFM).
The steady-state density is:
\[
e = \begin{bmatrix} 0.8443 &   0.8463 &   0.7956  &  0.9510  &  0.1814 &   0.1416   \end{bmatrix}',
\]
and the corresponding steady-state flow is $R=0.1557$.  Note that since~$\lambda_3$ is   high and~$\lambda_4$ is   low,  
$e_1,e_2,e_3,e_4 \gg e_5,e_6$, indicating a traffic jam at site $4$.  
Consider now the case $r=5$ (i.e. $q=1/5$). The steady-state density is now
\[
\tilde e = \begin{bmatrix} 0.5512  &  0.4645   & 0.2549  &  0.8969 &   0.0765 &   0.1963  \end{bmatrix}',
\]
and the corresponding steady-state output is $\tilde R=0.2820$. Note that now the density at site $4$ decreased relative to the $r=1$ case, and that $\tilde R>R$. Note also that $\sum_{i=1}^6 e_i = 3.7602 > \sum_{i=1}^6 \tilde e_i = 2.4403$. This means that the introduction of a 
``neighbor-repelling'' force (i.e.~$r>1$) alleviated the traffic jam, reduced the total steady-state occupancy,
 and increased the steady-state flow.

Fig.~\ref{fig:rfmi_n6_e} depicts the steady-state densities in this example 
 as a function of~$r\in[1,10]$. It may be 
observed that~$e_i$, $i=1,\dots,5$, monotonically decreases with $r$, and that $e_6$ slightly increases with $r$. Note that since the occupancy at site $6$ is not affected by $q$, but only by $r$, increasing $r$ should indeed increase $e_6$.~\hfill{$\square$}
\end{Example}

	\subsection*{Extreme interactions}

	To gain more insight on the effect of the nearest-neighbor 
	interactions on the steady-state behavior, it is useful to consider the cases when~$r \to  0$ 
	(so~$q=\frac{1}{r} \to \infty$)
	and~$r\to \infty$ (so~$q=\frac{1}{r} \to 0$).
	
	 \subsubsection*{The case~$r\to 0$} 

Intuitively speaking, a low value of~$r$   corresponds to:
 (1)~a strong attachment between existing 
 nearest neighbors (small~$r$);
and (2)~a high tendency for moving forward  if this involves  creating  new neighbors (large~$q$). 
As we will see  this leads to  the formation  of traffic jams and, consequently,
 to a sharp decrease in the output rate.

\begin{figure}[t]
  \begin{center}
  \includegraphics[width= 9cm,height=8cm]{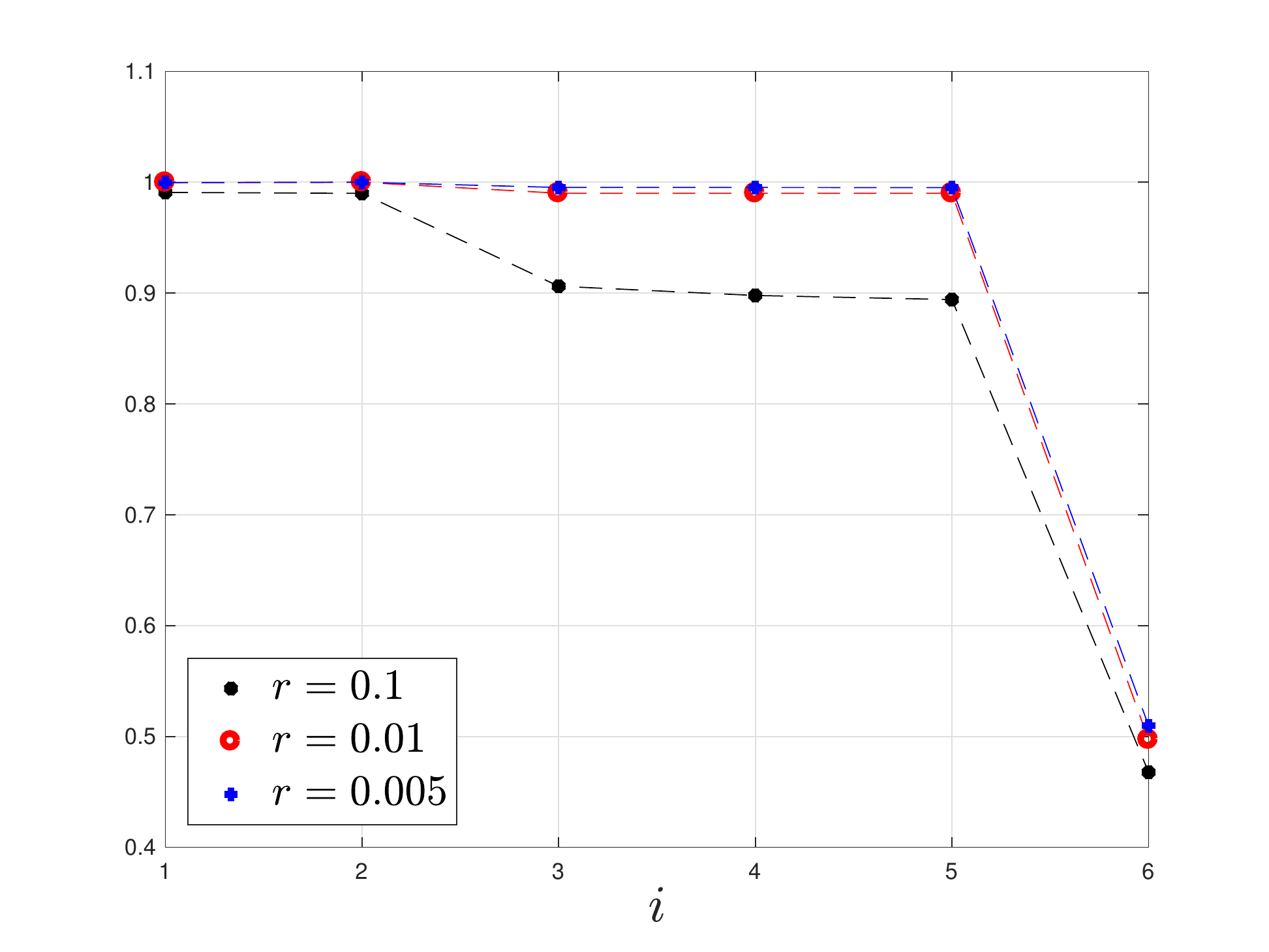}
 \caption{\updt{Steady-state densities~$e_i$ as a function of~$i$ for a~{\model} with~$n=6$, $\lambda_i=1$, $i=0,\dots,6$, for three values of $r$ (with~$q=1/r$).}}\label{fig:rfmi_n6_homg}
  \end{center}
\end{figure}

\begin{Example}\label{exp:rzero}
Consider a {\model} with dimension~$n=6$ and rates~$\lambda_i=1$, $i=0,\dots,6$. 
For $r=0.1$ (recall that $q=1/r$), the steady-state values are: 
\[
 e= \begin{bmatrix}  
     0.9908  &  0.9899 &   0.9062 & 0.8978 & 0.9841 & 0.5678
		\end{bmatrix}',\;\; R=    0.0913.
\]
For $r=0.01$, the steady-state values are: 
\[
  e=\begin{bmatrix}
    0.9998 &   0.9999 & 0.9901 & 0.9900 & 0.9899 & 0.4970
    \end{bmatrix}',\;\; R=    0.0099.
\]
For $r=0.005$, the steady-state density values are: 
\[
  e=\begin{bmatrix}
    0.9996 &   0.9999 & 0.9950 & 0.9950 & 0.9950 & 0.4986
    \end{bmatrix}',\;\; R=   0.0050.
\]
\updt{Fig.~\ref{fig:rfmi_n6_homg} depicts the steady-state values for the three~$r$ values.} It may be observed that
as~$r$ decreases the density in the first five sites  increases to one, i.e. these sites become completely full, and the output rate goes to zero. Note that this  highlights the negative
effect of   traffic jams on  the output rate.~\hfill{$\square$}
\end{Example}

We now rigorously analyze the case~$r\to 0$ for the~{\model} with~$n=2$ and~$n=3$. 
\begin{Example}\label{eq:case2}
Consider the~{\model} with~$n=2$ and~$q=1/r$. 
Expanding~$e_2$ and~$e_1$ in~\eqref{eq:expa2}
as a Taylor  series in~$r$ yields
\be \label{eq:ew234}
			e_2=1-\frac{\lambda_2}{\lambda_1}r+o(r),\quad e_1=1+o(r),
\ee
where every~$o(r)$ denotes a function~$f(r)$ satisfying~$\lim_{r\to 0} \frac{f(r)}{r}=0$. 
Thus,~$R=\lambda_1 e_1 (1-e_2)=\lambda_2 r+o(r)$. 
This implies in particular that 
\[
\lim_{r \to  0} e_1= \lim_{r \to 0}  e_2 =1,\quad \lim_{r \to 0} R=0.
\]
Thus, when~$r  \to  0$, both steady-state 
densities go to one,\footnote{Note that~\eqref{eq:ew234} implies that~$e_1$ goes to one faster than~$e_2$.} that is, the sites become completely full, 
and consequently 
the steady-state output rate goes to zero.~\hfill{$\square$}
\end{Example}


The next result analyzes the case~$n=3$.
\begin{Proposition}\label{eq:propen3}
The steady-state densities in the {\model} with~$n=3$ satisfy
\begin{align}\label{eq:rfm3ee}
e_1(r)&=  1-\frac{\lambda_2\lambda_3}{\lambda_0(\lambda_2+\lambda_3)} r^2+\operatorname{o}( r^2 ),  \nonumber \\
e_2(r)&=  1-\frac{\lambda_3}{\lambda_1} r^2+  \operatorname{o}( r^2 ) , \nonumber \\
e_3(r)&=   \frac{\lambda_2}{\lambda_2+\lambda_3} +\oo r  ,  
\end{align}
and
\be \label{eq:ren3}
R(r)=			\frac{\lambda_2\lambda_3 }{\lambda_2+\lambda_3} r+ \oo r 			.
\ee
\end{Proposition}
 Note that this implies that
\[
			\lim_{r\to 0} e_1(r)=\lim_{r\to 0} e_2(r)=1, \text{ and } \lim_{r\to 0} R(r)=0,
\]
so again as~$r\to 0$ sites at the beginning of the lattice become completely full and consequently
 the output rate goes to zero. 
 
Summarizing,  as $r$ goes to $0$  the repelling force between existing neighbors is very weak,
 and the binding force when forming new neighbors is very strong, leading to the formation of traffic jams at the beginning of the lattice.  
Consequently,  the steady-state flow   goes to zero.   

We now turn to consider the opposite case, that is,~$r\to \infty$.

	 \subsubsection*{ The case~$r\to \infty$} 

\begin{figure}[t]
  \begin{center}
  \includegraphics[width= 9cm,height=8cm]{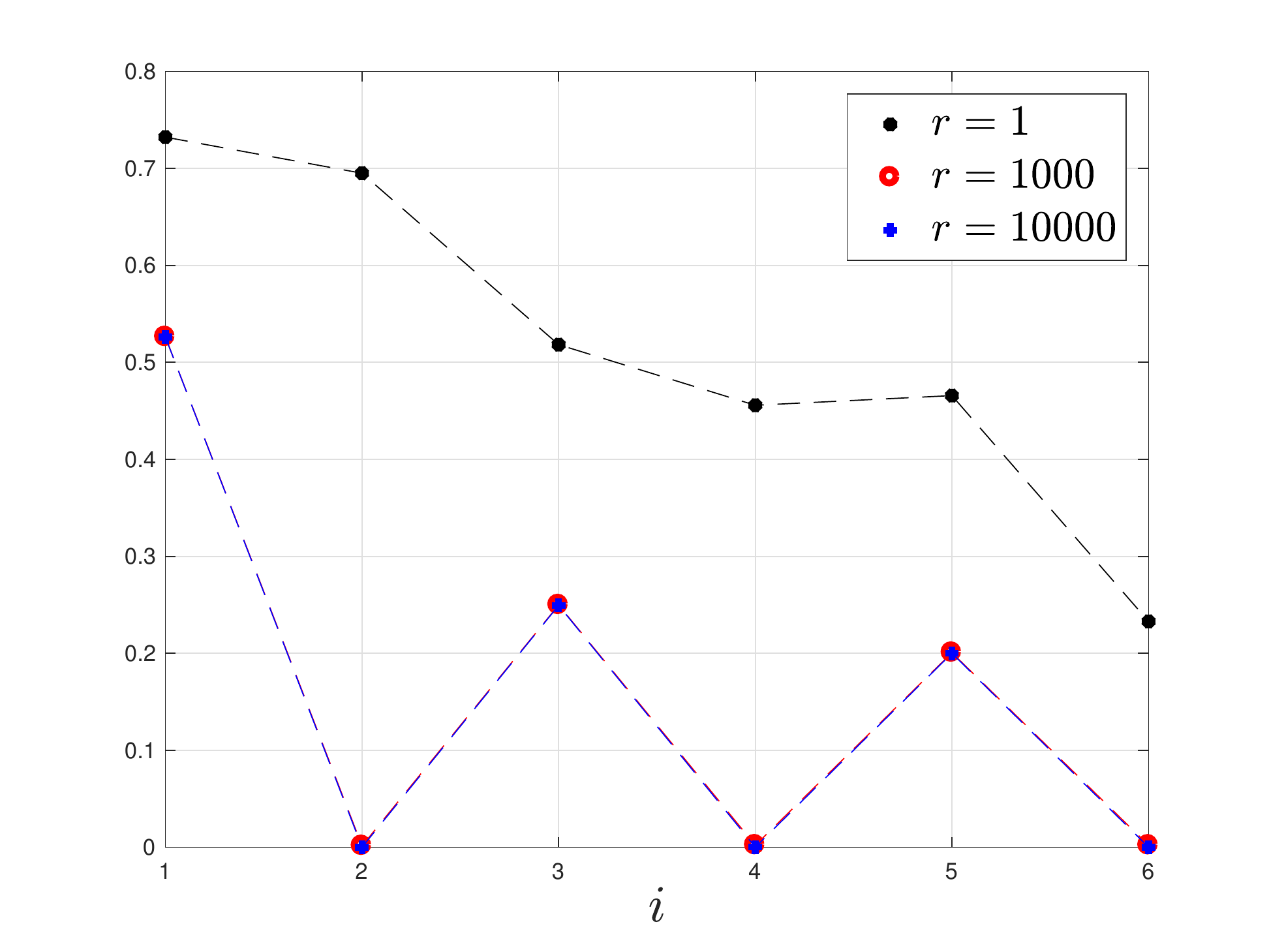}
 \caption{\updt{Steady-state densities~$e_i$ as a function of~$i$ for a~{\model} with~$n=6$, $\lambda_0=1,\lambda_1=1.2,\lambda_2=0.8,\lambda_3=0.95,\lambda_4=1.1,\lambda_5=0.75$, and $\lambda_6=1.15$, for three values of $r$, and~$q=1/r$. The steady-state values for $r=1,000$ and~$r=10,000$ cannot be  distinguished.}}\label{fig:rfmi_n6_r_large}
  \end{center}
\end{figure}

A  large  value of~$r$   corresponds to:
 (1)~strong repulsion  between existing 
 nearest neighbors (large~$r$);
and (2)~a low tendency for moving forward  if this involves  creating  new neighbors (small~$q$). 
As we will see  below, this leads to    a phenomena that may be regarded as the opposite of traffic jams, that is, a complete  
``separation of the densities'' along the lattice.  
\begin{Example}\label{exp:rinf}
Consider the {\model} with $n=6$ sites and rates
$
\lambda=\begin{bmatrix} 1 & 1.2 & 0.8 & 0.95 & 1.1 & 0.75 & 1.15     \end{bmatrix}'.
$
For $r=1$ (recall that $q=1/r$),  
\[
e= \begin{bmatrix}  0.7322  &  0.6950  &  0.5183  &  0.4558  &  0.4657  &  0.2329   \end{bmatrix}',\;\; R=0.2678.
\]
For $r=1,000$,  
\[
 e=\begin{bmatrix}  0.5262 &   0.0015 &   0.2498  &  0.0022   & 0.2007 &   0.0020   \end{bmatrix}',\;\; R=0.4729.
\]
For $r=10,000$,  
\[
 e=\begin{bmatrix}  0.5261  & 0.0001 &   0.2495 &   0.0002  &  0.2001 &   0.0002    \end{bmatrix}',\;\; R=0.4734.
\]
\updt{Fig.~\ref{fig:rfmi_n6_r_large} depicts these steady-state values for the three $r$ values. Note that the steady-state values for~$r=1,000$  and~$r=10,000$ cannot be distinguished.} It may be observed that the values~$e_j(r)$, $j=2,4,6$,  decrease to zero as~$r$ increases.
In other words, in every pair of consecutive sites  one density is very small. This ``separation of densities'' represents the opposite of a traffic jam. 
This leads to a substantial 
 increase in  the output rate~$R$  as~$r$ increases.~\hfill{$\square$}
\end{Example}

We now rigorously analyze the case~$r\to \infty$ for the~{\model} with~$n=2$ and~$n=3$. 
\begin{Example}\label{eq:case2qq}
Consider the~{\model} with~$n=2$. 
Expanding~$e $ in~\eqref{eq:expa2}
as a Taylor  series in~$q=1/r$ yields
\[
			e_2=\frac{\lambda_1}{\lambda_2}q+o(q),\quad e_1=\frac{\lambda_0}{\lambda_0+\lambda_1}+o(q),
\]
so 
\[
\lim_{r \to \infty} e_2=0 ,\;\; \lim_{r \to \infty }  e_1 =\frac{\lambda_0}{\lambda_0+\lambda_1}, \;\;\lim_{r \to \infty}  R = \frac{\lambda_0\lambda_1}{\lambda_0+\lambda_1}.
\]
Thus, in this case the density at site~$2$ goes to zero, 
and   this   yields   a positive
steady-state output rate.~\hfill{$\square$}
\end{Example}

\begin{Proposition}\label{eq:qq3}
The steady-state densities in the {\model} with~$n=3$ satisfy
\begin{align}\label{eq:rfm3qq}
e_1(q)&=  a_1  +b_1 q + \oo q ,  \nonumber \\
e_2(q)&=  \frac{\lambda_1}{\lambda_2} q+ \oo q  , \nonumber \\
e_3(q)&=  a_3+b_3 q  +\oo q  ,  
\end{align}
with~$a_1,a_3 \in (0,1)$, and
\be \label{eq:ren3qq}
R(q)=			\lambda_0 (1-a_1 ) +	\lambda_0  \left ( (a_1-1) \frac{\lambda_1}{\lambda_2 } -b_1 \right  ) q +\oo q 		.
\ee
\end{Proposition}

 Note that this implies that
\[
			 \lim_{r\to \infty} e_2(r)=0, \text{ and } \lim_{r\to \infty} R(r)>0,
\]
so again as~$r\to \infty$ the density at site~$2$ goes to zero and the output rate  is positive.

\section*{Discussion}
Motor proteins and other moving biological particles interact with their neighbors. 
\updt{Indeed, it is known that 
cellular cargoes are often moved by \emph{groups} of motor proteins, and
 recent findings suggest 
that the bounding time of kinesins on microtubules  
  depend on the
presence of neighbors.
}

To study the effect of such interactions, 
we introduced a new deterministic
compartmental  model, the~{\model}, for the flow of particles along an ordered
 lattice of sites where the
transition rates  between sites depend both on properties of the lattice and  on 
   nearest-neighbor interactions between the particles. 
\updt{The properties of the lattice are modeled using transition rates~$\lambda_i$ between sites.
The nearest-neighbor interactions between the particles are modeled using two parameters:~$r$  that represents
the tendency of a moving particle to  break from  an existing neighbor,
 and~$q$  that represents
the tendency of a   particle to move into a site   such that   it forms new neighbors (see Fig.~\ref{fig:rfmi_sitei}). 

}

\updt{The {\model} is based on a mean-field ansatz   neglecting high-order 
 correlations of occupations between neighboring sites. 
It is possible to use our framework also to 
derive a more complete model based on  binary occupation densities and transitions 
described by a continuous-time  master equation (see, e.g. the interesting paper~\cite{bauer_nadler_2013} 
in which this was done for granular channel transport). However, in such a 
model the state-variables at time~$t$ represent the probability of each configuration at time~$t$,
and the number of possible configurations grows exponentially   with the number of sites~$n$. 
On the other-hand,  the~{\model} includes~$n$ (nonlinear) ODEs for~$n$ sites.
 Another  important advantage of the~{\model}  is that
     it is   amenable to analysis  using tools from systems and control theory, even in the non-homogeneous case. This allows to rigorously study, for example, the effect of the nearest-neighbors interactions on the steady-state behavior of the~{\model}  for any set of transition rates. Our results show that suitable forces between nearby particles can greatly increase the
output rate, and reveal that the underlying mechanism for this is the alleviation of traffic jams along the lattice. In particular, when
the parameter~$r$ is very large and~$q$ is very small, the steady-state density
is such that any second site is empty. This represents the ``opposite'' of  a traffic jam, and increases 
 the steady-state flow.}

The phenomenological model introduced here may   prove   useful for other applications as well. 
For example, an important problem in   vehicular traffic  is
 to understand how human drivers
react to  nearby cars. 
One may also consider  implementing   appropriate nearest-neighbor dynamics
in algorithms that control autonomous vehicles in order to reduce  traffic jams and increase the flow.
  Of course, implementing this with a very large~$r$ (or~$q$)
means very high effective transition rates, but our results suggest that even for~$r$ not much larger than one the increase in the flow is
non-negligible.  Another interesting topic for further research is generalizing the~{\model} to include the possibility of
attachment/detachment of particles from intermediate sites in the lattice (see~\cite{brfm} for some related ideas).

\section*{Acknowledgments}
We thank the anonymous referees and the Associate Editor for their helpful comments.

The research of~YZ is partially supported by the Edmond J. Safra Center for Bioinformatics at Tel Aviv University. The research of MM is partially supported by research grants from the Israeli Ministry of Science, Technology \& Space, the US-Israel  Binational Science Foundation, and  the Israeli Science Foundation.

\section*{Appendix: Proofs}
{\sl Proof of Proposition~\ref{prop:inv}.}
  The fact that~$C^n$ is an invariant set of the dynamics follows immediately from the equations of the {\model}. Let
\be\label{eq:eta}
  \eta_i(t):=\lambda_i(1+(q-1)z_{i+2}(t))(1+(r-1)z_{i-1}(t)), \quad i=0,\dots,n,
 \ee
with the~$z_i$s defined in~\eqref{eq:defz}.
 By~\eqref{eq:rfmi_odes}, the {\model} can be written as 
 \be\label{eq:rfmi_rfm}
 \dot{x}_i(t)=\eta_{i-1}(t)x_{i-1}(t)(1-x_i(t))-\eta_i(t) x_i(t)(1-x_{i+1}(t)).
 \ee
This is just the RFM (see~\eqref{eq:rfm_all}), but  with time-varying 
rates~$\eta_i(t)$. 
Let~$a_i:= \min\{1,q\} \min\{1,r\}\lambda_i$, and~$b_i:= \max\{1,q\} \max\{1,r\}\lambda_i$.
It follows from~\eqref{eq:eta}
that~$a_i \leq \eta_i(t)\leq b_i$ for all~$i$ and for
all~$t\geq 0$. Note that for~$r,q>0$ every~$a_i$ is strictly positive.
In other words, all the time-varying
rates are uniformly separated from zero and uniformly bounded. 
Now the   proof  of   Proposition~\ref{prop:inv}
follows from the results in~\cite{RFM_entrain}.~\IEEEQED

{\sl Proof of Proposition~\ref{prop:weak_cont}.}
  Combining the representation in~\eqref{eq:rfmi_rfm} with the uniform 
	boundedness of the rates, Proposition~\ref{prop:inv},
	and 
  the results in \cite{3gen_cont_automatica} imply that 
the {\model} is  \emph{contractive after a small overshoot and short transient}~(SOST) on~$C^n$.
Also,  Proposition~4 in~\cite{3gen_cont_automatica} implies that for the {\model} the properties
of~SOST and~SO are equivalent, and this completes the proof.~\IEEEQED

	{\sl Proof of Fact~\ref{fact:case2simp}.}
	Consider the~{\model} with~$n=2$ and~$q=1/r$. Then~\eqref{eq:eq_of_model} becomes
\begin{align*} 
										    \lambda_0   (1-e_1)  (1+(\frac{1}{r}-1)e_2)          
										 & = \lambda_1 e_1(1-e_2)             \\
										  & =  \lambda_2  e_2  (1+(r-1)e_1) .
\end{align*}
This yields 
\be\label{eq:e1case2}
			e_1=\frac{\lambda_2 e_2}{\lambda_1+( \lambda_2(1-r)-\lambda_1  )e_2 }
\ee
and 
\[
				a_1 e_2^ 2 +a_2 e_2+\lambda_1 =0,
\]
with~$a_1$ defined in~\eqref{eq:defais}
 and~$a_2 :=-\lambda_1-\lambda_2-a_1$.  
The feasible solution (i.e. the one satisfying~$e_1,e_2\in(0,1)$ for any set of parameter values) is
given by 
\[
e_2=\frac{  -a_2 - \sqrt{ a_2^2-4 a_1 \lambda_1  }   }{2 a_1},
\]
and~\eqref{eq:e1case2}.~\IEEEQED
	
{\sl Proof of Prop.~\ref{prop:ediff}.} 
To emphasize the dependence on the parameters, write the {\model} as~$\dot x=f(x;v)$,
where~$  v:=\begin{bmatrix}
\lambda_0& \dots& \lambda_n & r & q \end{bmatrix}'$. Note that~$f$ is an analytic function. 
Then the steady-state satisfies the relation~$f(e;v)=0$. The Jacobian matrix of
this relation with respect to~$x$ is
\[
J(x;v):=\frac {\partial }{\partial x} f(x;v),
\]
which is just the Jacobian of the dynamics. Fix~$v^0\in \R^{n+3}_{++}$ and let~$e^0\in\Int(C^n)$
denote the corresponding steady-state, that is,~$f(e^0;v^0)=0$ and~$e^0=h(v^0)$. 
Suppose that there exists a matrix measure~$\mu:\R^{n\times n}\to \R$ such that~$\mu( J(e^0;v^0 ) )<0$.
This implies in particular that~$J(e^0,v^0)$ is Hurwitz (see e.g.~\cite{Desoer_cont}), so it is not singular and invoking
the implicit function theorem implies that the  mapping~$h$ is analytic. 
It follows from the results in~\cite{RFM_entrain} that such a matrix measure~$\mu$
 indeed exists, and this
completes the proof.~\IEEEQED

{\sl Proof of Prop.~\ref{eq:propen3}.}
Expand~$e_i$, $i=1,2,3$, as
\be\label{eq:expeip}
e_i=a_i+b_i r+c_i r^2 +\operatorname{o}( r^2 ).
\ee
Recall that the steady-state equations are given by~$R(e)=g_0(e)=g_1(e)=\dots=g_3(e)$, 
with the~$g_i$s given  in~\eqref{eq:rfmi_g}. Substituting~\eqref{eq:expeip}   yields
\begin{align*}
			g_0(e)&=\frac{\lambda_0(1-a_1)a_2}{r}+\dots,\\
			g_1( e)&=\frac{\lambda_1 (1-a_2)a_3}{r}+\dots,\\
			g_2( e)&=\lambda_2 (a_1-1)a_2(a_3-1) +\dots,\\
			g_3( e)&=\lambda_3 (1-a_2)a_3 +\dots .
\end{align*}
Since~$R(e)$ is bounded, we conclude  that
\be\label{eq:a3a433}
(1-a_1)a_2=(1-a_2)a_3=0. 
\ee
Assume for the moment that~$a_2=0$. Then
\begin{align}\label{eq:poitr}
			g_2( e)&=\lambda_2 (a_1-1) (a_3-1)b_2 r +\dots,\\
			g_3( e)&=\lambda_3  a_3 +\dots ,\nonumber
\end{align}
and this implies that~$a_3=0$. Now, $g_1(e)=\lambda_1 (1+b_3)+\dots$ and combining this with~\eqref{eq:poitr}
yields~$b_3=-1$. Thus, $e_3=a_3+b_3 r+c_3 r^2+\operatorname{o}( r^2 )=-r+\oo r$,
 and this is a contradiction as~$e_3(r)$ will be strictly negative  for any~$r>0$ sufficiently small. 
  We conclude that~$a_2\not =0$, so~\eqref{eq:a3a433} yields~$a_1=1$,
 and also~$(1-a_2)a_3=0$. 
Suppose that~$a_3=0$. Then 
\begin{align*}
			g_0(e)&=-\lambda_0 a_2 b_1+\dots,\\
			g_1( e)&=\lambda_1(1+b_3)(1-a_2)+\dots,\\
			g_2( e)&=\lambda_2 a_2(b_1-1)r +\dots,\\
			g_3( e)&=\lambda_3(1-a_2) b_3 r +\dots .
\end{align*}
It follows that~$a_2 b_1= (1+b_3)(1-a_2)   =0 $. Since we already know that~$a_2\not =0$, $b_1=0$.
The case~$b_3=-1$ is impossible, as then~$e_3(r)<0$ for~$r>0$ sufficiently  small, so~$a_2=1$. 
But then~$R(e)=g_2(e)=- \lambda_2   r +\dots$ and this is a contradiction. We conclude that~$a_3\not =0$, so~\eqref{eq:a3a433} yields~$a_2=1$. Summarizing, we have~$a_1=a_2=1$. Now,
\begin{align*}
			g_0(e)&=-\lambda_0   b_1+\dots,\\
			g_1( e)&=-\lambda_1 b_2 a_3+\dots,\\
			g_2( e)&=\lambda_2 (a_3-1) (b_1-1)r +\dots,\\
			g_3( e)&=\lambda_3(1-b_2) a_3 r +\dots .
\end{align*}
This gives~$b_1=0$ and~$b_2 a_3=0$. Since we already know that~$a_3\not =0$, $b_2=0$. Now,
\begin{align*}
			g_0(e)&=-\lambda_0  c_1 r+\dots,\\
			g_1( e)&=-\lambda_1   a_3 c_2 r+\dots,\\
			g_2( e)&=\lambda_2 (1-a_3)  r +\dots,\\
			g_3( e)&=\lambda_3  a_3 r +\dots .
\end{align*}
Equating the coefficients here yields~$a_3=\frac{\lambda_2}{\lambda_2+\lambda_3}$, 
$c_1=\frac{-\lambda_2\lambda_3} {\lambda_0(\lambda_2+\lambda_3)}$, and~$c_2=-\lambda_3/\lambda_1$.
Since we know that the steady-state equations admit a unique solution this yields~\eqref{eq:rfm3ee}, and
the equation~$R(e)=g_0(e)$ yields~\eqref{eq:ren3}.~\IEEEQED

{\sl Proof of  Prop.~\ref{eq:qq3}.}
 Expand~$e_i$, $i=1,2,3$, as
\be\label{eq:expeiqq}
e_i=a_i+b_i q+c_i q^2 +\operatorname{o}( q^2 ).
\ee
Recall that the steady-state equations are given by~$R(e)=g_0(e)=g_1(e)=\dots=g_3(e)$, 
with the~$g_i$s given  in~\eqref{eq:rfmi_g}. Substituting~\eqref{eq:expeiqq}   yields
\begin{align*}
			g_0(e)&=\lambda_0 (a_1-1) (a_2-1)+\dots,\\
			g_1( e)&=\lambda_1 a_1(a_2-1)(a_3-1) +\dots,\\
			g_2( e)&=\lambda_2 \frac{ a_1a_2(1-a_3) }{q} +\dots,\\
			g_3( e)&=\lambda_3 \frac{  a_2 a_3  }{q}  +\dots .
\end{align*}
This implies  that
\be\label{eq:a3aqqq}
a_1a_2(1-a_3) = a_2a_3=0. 
\ee
Assume for the moment that~$a_2\not = 0$. Then~$a_3=a_1=0$. This yields  
\begin{align*}
			g_0( e)&=\lambda_0 (1-a_2) +\dots,\\
g_1( e)&=\lambda_1 (1-a_2) b_1 q +\dots,\\
			g_2( e)&=\lambda_2  a_2(1+b_1) +\dots. 
	\end{align*}
	This implies that~$a_2=1$ and~$b_1=-1$. This yields~$e_1(r)=-r+\oo r$ which is a contradiction.
	
	We conclude that~$a_2=0$. Now, 
	\begin{align*}
			g_0(e)&=\lambda_0 (1-a_1 )  +   \dots,\\
			g_1( e)&=\lambda_1 a_1 (1-a_3) +   \dots,\\
			g_2( e)&=\lambda_2 a_1(1-a_3)b_2  +    \dots,\\
			g_3( e)&=\lambda_3  a_3 (1+b_2)  +     \dots .
\end{align*}
Equating the coefficients here yields  the following. First,~$a_1\not =0$, and since~$e_1(q)=a_1+\dots$, this implies that~$a_1>0$. 
Second, if~$a_3=1$ then~$a_1=1$ and~$b_2=-1$ which is a contradiction as then~$e_2(q)=-q+o(q)$.
Thus,~$a_3 \not =1$ and this implies that~$a_1 \not = 1$.  
We conclude that~$a_1,a_3 \in (0,1)$. Now the equations for~$g_1$ and~$g_2$ yield~$b_2=\lambda_1/\lambda_2 $.
This proves~\eqref{eq:rfm3qq}. Expanding~$g_0 $ up to order one in~$q$, and using~$R=g_0$ yields~\eqref{eq:ren3qq}.~\IEEEQED

 

\end{document}